\documentclass[12pt]{book}

\usepackage[dvips]{graphicx,color}
\usepackage{makeidx,tsukuba}

\makeauthorindex
\makeindex

\begin{document}

\BookTitle{\itshape The 28th International Cosmic Ray Conference}
\CopyRight{\copyright 2003 by Universal Academy Press, Inc.}
\pagenumbering{arabic}

\chapter{
Numerical Likelihood Analysis of Cosmic Ray Anisotropies
}

\author{%
%
%
Carlos Hojvat,$^1$ Thomas P. McCauley,$^2$  
Stephen  Reucroft,$^2$ and John D. Swain$^2$\\
{\it 
(1) Fermi National Accelerator Laboratory, P.O. Box 500, Batavia, IL 60510 \\
(2) Department of Physics, Northeastern University, Boston, MA 02115}
}

\section*{Abstract}
A numerical likelihood approach to the determination of cosmic ray
anisotropies is presented which offers many advantages over other
approaches. It allows a wide range of statistically meaningful
hypotheses to be compared even when full sky coverage
is unavailable, can be readily extended in order to include
measurement errors, and makes maximum unbiased use of all
available information.

%
%

\section{Introduction}

The search for anisotropies in the cosmic ray distribution (and of other
objects -- see for example~[4,5]) is of increasing
interest with the advent of the Pierre Auger Observatory (for a review of
the physics up to this time, see [1]). The issues involved are, however, subtle
and complicated by limited statistics at the highest energies and nonuniform
sky coverage. The first full-sky search for UHECR anisotropies using a standard approach
is presented in this conference~[2]. 
This paper presents a maximum likelihood approach which
is well-adapted to further studies in anticipation of much larger data sets.

\section{Expansion in Spherical Harmonics}

It is convenient to define a set of real spherical harmonics
$\lbrace \psi_{\ell,m} \rbrace$ for $\ell=0,1,2,\ldots$ and 
$m=-\ell,-(\ell-1),\ldots,0,\dots,(\ell-1),\ell$
by
\begin{equation}
\psi_{\ell,m}(\theta,\phi) = \left\{ 
\begin{array}{ll}
k_\ell^{|m|}P_\ell^{|m|}(\cos\theta)\cos(m\phi) & \mbox{for\  $m=-\ell,\ldots,-1$},\\
k_\ell^0P_n^0(\cos\theta) & \mbox{for\  $m=0$,}\\
k_\ell^mP_\ell^{m}(\cos\theta)\sin(m\phi) & \mbox{for\  $m=1,\ldots,\ell$}
\end{array} \right. 
\end{equation}
\noindent
which are orthonormal with respect to the usual measure
$\sin(\theta)d\theta d\phi$ and integration over $\theta$ from 0 to $\pi$
and $\phi$ from 0 to $2\pi$, the $P_\ell^m$ are associated Legendre
polynomials of the first kind, and the normalization constants $k^\ell_m$ are:
\begin{equation}
k_0^0=\frac{1}{\sqrt{2\pi}}, k_\ell^0=\sqrt{\frac{2\ell+1}{4\pi}}, {\mathrm{\ and\ }}
k_\ell^m=\sqrt{\frac{2\ell+1}{2\pi}\frac{(\ell-m)!}{(\ell+m)!}}
\end{equation}
\noindent
The natural measure of anisotropy for a spherical 
distribution
is in terms of these spherical harmonics, as each $\ell$ labels, 
in a coordinate-independent fashion, just how much of each irreducible
$SO(3)$ representation is present.

In a perfect world with infinite statistics and complete sky coverage
there are now many possible approaches to estimating how much of each
of these components is present in a distribution, or, better,
what is the likelihood that a given function 
$f(\theta,\phi)$ with Fourier-Legendre expansion
$f(\theta,\phi)=\sum_{\ell=0}^\infty \sum_{m=-\ell}^{\ell}a_{\ell,m}\psi_{\ell,m}$ representing
the probability density of sources gives rise to the observed distribution
$g(\theta,\phi)$. The coefficients can be extracted from the usual
integral 
\begin{equation}
a_{\ell,m} = \int_0^{2\pi} \int_0^\pi f(\theta,\phi)\psi_{\ell,m}(\theta,\phi)\sin\theta d\theta d\phi
\end{equation}
\noindent
Of course this gives no measure of what sort of error should be associated with 
the determined values of each coefficient.
Alternative
approaches are to fit for the coefficients by minimizing some $\chi^2$-like
quantity like
\begin{equation}
\int_0^{2\pi} \int_0^\pi \frac{\left|g(\theta,\phi)- 
\sum_{\ell=0}^\infty \sum_{m=-\ell}^{\ell}a_{\ell,m}\psi_{\ell,m}(\theta,\phi)\right|^2}{\sigma^2}
\sin\theta d\theta d\phi
\end{equation}
\noindent
with $\sigma$ a suitable measure of error, 
or to compute and maximize a corresponding likelihood that the hypothesized
distribution parametrized by the $a_{\ell,m}$ gives rise to the the observed
distribution $g$.
Of course this is statistically unreasonable for a finite number of sources
({\em i.e.} to provide an infinite number of coefficients!), 
In general a decision must be made to truncate the expansion
at some value of $\ell$ to make the sum finite, but now the general phenomenon
of aliasing risks that, for example, a fit allowing for $\ell=0,1$ might
give misleading results for observed data which is drawn from a purely
$\ell=2$ distribution, say. 

A more serious problem is that should part of the sky be unobserved, there is now
no way to calculate anything! This is not a trivial point. An attempt to find 
anisotropies based only on observations in the Northern hemisphere with zeroes
inserted for the whole Southern hemisphere would be wildly
in error if some simple extrapolation were made to the unobserved region of the sky  --
especially if there were something bright and as-yet undetected in the South!
The real challenge is to say something statistically meaningful with the data
that is actually available. Clearly, observing only the Northern hemisphere and seeing
a good degree of isotropy should increase one's net belief in overall isotropy 
of the full sky, while leaving open the possibility of a staggeringly bright or
empty sky in the South.
One approach is to try to make functions which would be orthonormal over the
observed region of the sky, but it's not clear that this has much physical meaning as
it elevates lack of acceptance to a status comparable to the $SO(3)$ invariance
of space embodied in the $\psi_{\ell,m}$. The following is a proposal for what
seems to make good statistical sense.

\section{A Likelihood Proposal}

Based on the above observations, we make the following proposal: keep the spherical
harmonics as always with the (necessarily) truncated Fourier-Legendre expansion and construct an
unbinned likelihood~[3] function in which one clearly specifies which values of $\ell$
are included in the sum. An unbinned likelihood function, as described below, automatically
makes maximum use of all detected information, allows for measurement errors to
be included easily, and is easy to implement numerically. Most importantly, however,
the likelihood is {\em not} to be normalized as it stands. Rather, we take the Bayesian approach
to likelihood which says that likelihoods give us ways to update our prior experimental
data or guesses in light of new information. This will mean that we are able to present
results on various hypotheses about data without bias, and with the easy inclusion
of other data from the same, or other experiments. 

To be concrete, we specialize here to the case where $g$ is a sum of
delta functions representing sources $i$ of unit intensity at $\theta_i,\phi_i$
and later discuss how one can treat the case of these being at uncertain
locations, or taking into account other properties such as intensity,
energy, composition, {\em etc.}. These will appear as natural generalizations
to the approach described.

The (unnormalized!) likelihood $L(\lbrace \ell\rbrace|\theta_i,\phi_i)$ that the measured $\theta_i,\phi_i$ arise
from $f(\theta,\phi)=\sum_{\ell}^\infty \sum_{m=-\ell}^{\ell}a_{\ell,m}\psi_{\ell,m}$
where the sum over $\ell$ is specified according to whatever hypothesis is being
tested ({\em i.e.} just taking $\lbrace\ell\rbrace=\lbrace 0,1\rbrace$ allows for uniform and dipole contributions
and no others, while $\lbrace\ell\rbrace=\lbrace2\rbrace$ would be pure quadrupole) is
\begin{equation}
L(\lbrace \ell\rbrace|\theta_i,\phi_i) = 
\frac{f(\theta_i,\phi_i)}{\int \prod_{{\ell}}\prod_{m=-\ell}^{\ell} da_{\ell,m} f(\theta,\phi) }
\end{equation}
\noindent
where the integral in the denominator is over all the parameters that can vary and over the
range in which the parameters are allowed to vary. An important caveat is that for $f$ to represent
a sensible probability distribution it should never be negative, and this must be checked.
Two approaches are possible: one is to restrict the domain over which coefficients range so
that the function is strictly positive (this is not actually very difficult in practice
since distributions are often nearly uniform with small fluctuations superimposed) or to 
take as a probability distribution function some positive function of $f$ in place of $f$
above. In astronomy~[4]
it is not uncommon to see $\exp(f)$. The choice ultimately represents the unavoidable
presence of some (often hidden) assumption about what a sensible prior ({\em i.e.}
in the absence of data) is for the likelihood -- a point to which we return later.

By construction then $L(\lbrace \ell\rbrace|\theta_i,\phi_i)$ is normalized so that its integral 
over all the parameters that can vary (the coefficients $a_{\ell,m}$ included in the truncated
Fourier-Legende expansion, and the total likelihood $L_{TOT}$, which is a function of those $a_{\ell,m}$ is
\begin{equation}
L_{TOT}(\lbrace \ell\rbrace)= \prod_i L(\lbrace \ell\rbrace|\theta_i,\phi_i)
\end{equation}
\noindent
This is a relative likelihood, and while no absolute normalization is possible (nor should it be!)
if part of the sky is unobserved, it is now very useful for two types of calculation. If one has
a prior expectation for the distribution of the $a_{\ell,m}$ (which might be that they are all
a priori equally likely) then $L_{TOT}$ can be multiplied by this and 
the product treated as a
normalized likelihood distribution
for the $a_{\ell,m}$ themselves. This is, of course, potentially dangerous, but does allow one
to see how the new data (the measured $(\theta_i,\phi_i)$) should cause one to revise earlier
beliefs. Such a likelihood can be maximized with respect to the $a_{\ell,m}$ and
if not enough data is present to test the hypothesis (for example, more 
coefficents
to fit for than data points) the fit will respond by just not converging 
(that is to
say, there will be flat directions in the likelihood as a function of the 
parameters
meaning that one can't decide) - the beauty of this approach is that it 
is, by
construction, correct.
When results are obtained, one automatically gets the values of the 
parameters, and  their entire likelihood distributions 
from which errors
(which need
not even  be Gaussian) can be extracted. One can even do things like fix some parameters 
to those given by a favoured theory and then repeat the fitting and 
then obtain likelihoods for the correctness of that theory.

More objectively, one can compute relative likelihoods in which the prior drops out,
so that it is reasonable to ask (even in the absence of full sky coverage!) what the relative
likelihood $L_{REL}$ is of pure dipole distribution to one admitting uniform, dipole and quadrupole
components:
%
$L_{REL} = \frac{L_{TOT}(\lbrace 1\rbrace)}{L_{TOT}(\lbrace 0,1,2\rbrace)}$
%
If one wants statements made to be about all energies over $x$, then one
just uses the
data points with energies over $x$. Similarly, data can be selected by
composition and (even relative) anisotropies be searched for as functions
of composition, energy, time, {\em etc.}
Extensions to uncertainty in direction are trivial to include: simply
divide a given event into
a large number, $M$, say of subevents  distributed appropriately and count 
each in the likelihood with
a weight $1/M$ -- this numerically convolves this uncertainty with the 
likelihood, and the size of any additional errors introduced by the procedure
can be studied by varying $M$.

\section{References}


\re
1.\ Anchordoqui L., Paul T., Reucroft S., Swain J., 2002
arXiv:hep-ph/0206072.
\re
2. Anchordoqui, L., Hojvat, C., McCauley, T. P.,Paul, T.C., Reucroft, S., Swain, J. D., Widom, A.,
these proceedings and arXiv:astro-ph/0305158.
\re
3. Edwards, A. W. F. 1972. Likelihood. Cambridge University Press, Cambridge.
\re
4. Evans, N. W., Ferrer, F., Sarkar, S., 
Astropart. Phys. {\bf 17} (2002) 319. 
\re
5. Peebles, P. J. E., Ap. J. {\bf 185} (1973) 413;
Hauser, M.G., Peebles P. J. E., Ap. J. {\bf 185} (1973) 757.
\endofpaper
\end{document}